\documentclass[aps,pra,pdf,superscriptaddress,amsmath,amssymb,amsfonts,twocolumn,showpacs,nofootinbib]{revtex4-2}

\usepackage{amssymb}
\usepackage{eqnarray,amsmath}
 
\usepackage{graphicx}
\usepackage{epsfig}
\usepackage{dcolumn}
\usepackage{bm}
\usepackage{braket}
\usepackage{amsmath}
\usepackage{physics}
\usepackage{csquotes}
\usepackage{mathtools}
\usepackage{graphicx,color,xcolor,colortbl}

\usepackage[colorlinks=true,
linkcolor=blue,
filecolor=blue,      
urlcolor=blue,
citecolor=blue]{hyperref}

\usepackage[normalem]{ulem}

\definecolor{Gray}{gray}{0.85}
\definecolor{LightCyan}{rgb}{0.88,1,1}

\newcolumntype{a}{>{\columncolor{Gray}}c}
\begin{document}

\title{Soliton-to-droplet crossover in a dipolar Bose gas in one and two dimensions}

\author{M. Schubert}
\email{malte.schubert@fysik.lu.se}
\affiliation{Mathematical Physics and NanoLund, Lund University, Box 118, 22100 Lund, Sweden}

\author{T. Bland}
\affiliation{Mathematical Physics and NanoLund, Lund University, Box 118, 22100 Lund, Sweden}
\affiliation{Institut für Experimentalphysik, Universität Innsbruck, Innsbruck 6020, Austria}

\author{M. J. Mark}
\affiliation{Institut für Quantenoptik und Quanteninformation, Österreichische Akademie der Wissenschaften, Innsbruck 6020, Austria}
\affiliation{Institut für Experimentalphysik, Universität Innsbruck, Innsbruck 6020, Austria}

\author{F. Ferlaino}
\affiliation{Institut für Quantenoptik und Quanteninformation, Österreichische Akademie der Wissenschaften, Innsbruck 6020, Austria}
\affiliation{Institut für Experimentalphysik, Universität Innsbruck, Innsbruck 6020, Austria}

\author{S. M. Reimann}
\affiliation{Mathematical Physics and NanoLund, Lund University, Box 118, 22100 Lund, Sweden}

\begin{abstract}
We analyze a system of dipolar atoms confined in geometries of quasi-low-dimensionality. Due to the long-range and anisotropic nature of dipolar interactions, the system supports both stable solitons and quantum droplets. In quasi-one-dimensional geometries, the transition between these states is known to manifest either as a first-order phase transition, associated with bistability, or as a smooth crossover. We investigate this transition by calculating the structure factor and showing that the response of the breathing mode provides an experimentally accessible probe. In addition, we identify regions of both bistability and smooth crossover in quasi-two-dimensional geometries. Finally, we connect our findings to previous experimental results and delineate the conditions under which two-dimensional dipolar bright solitons can be realized.
\end{abstract}
\date{\today}
 \maketitle
\section{Introduction}
Self-binding occurs across many areas of physics, from giant galaxies~\cite{Sivaram2021} and stars~\cite{Tielens2005} to water droplets~\cite{Young1805}, atoms~\cite{Sutton1988}, nuclei~\cite{gamow1930}, and optics~\cite{Duree1993}. In each case, competing energy contributions lead to structures of finite size. A necessary condition for self-bound states is the presence of nonlinearity in the governing equations of motion. For solitons, the balance between nonlinearity and dispersion~\cite{Drazin1989} results in a stable, non-dispersive wave~\cite{Zakharov1973,Zakharov1979,Rajaraman1982,Emplit1987}. Droplets, in contrast, achieve stability through a balance between surface tension and internal pressure~\cite{Bossert2023} and can remain stable in free space, unlike solitons. Moreover, the size of a soliton is restricted due to the increasing nonlinearity when scaling up the system~\cite{Chiao1964,Gammal2001,Salasnich2007}, whereas the droplet requires a minimum system size to overcome the internal pressure~\cite{Strutt1879,Petrov2015a}.

Bose-Einstein condensates (BECs) provide a versatile platform to study both solitons and droplets. Bright solitons can form in attractive BECs~\cite{Strecker2002,Cornish2006,Lepoutre2016}, and may exist in free space or two dimensions (2D) under certain conditions~\cite{Zhang2015,Saito2003,Wang2020,Chen2021}, but they are generally more stable in one dimension (1D)~\cite{Wen2016,Barbiero2014,DiCarli2019}. Early on, it was suggested that self-bound complexes, stabilized by higher-order terms beyond mean field, can form in bosonic multicomponent systems \cite{Bulgac2002,Petrov2015a,Kadau2016a}. Quantum droplets, for example, can form in atomic Bose mixtures~\cite{Cabrera2018,Semeghini2018,Cavicchioli2025,d2019observation,Guo2021}. By carefully tuning the interactions, the total mean-field interaction can be canceled out~\cite{Petrov2015a}. Their behavior was also studied in one~\cite{Astrakharchik2018,Petrov2016} and two dimensions~\cite{Spada2024,Dong2022}. In Bose-Bose systems, quantum droplets only exist due to a delicate balance between interactions and atom number, which makes them challenging to produce and often short-lived~\cite{Petrov2015a}.

Cheiney {\it et al.}~\cite{cheiney2018bright} investigated the occurrence of soliton and droplet states in a non-dipolar mixture of two spin hyperfine states of $^{39}\rm K$. Their study focuses on the transition between bright solitons and droplets, which can manifest either as a smooth crossover or as a first-order phase transition, depending on the interaction strengths. In the case of a first-order transition, bistability arises: both the soliton and droplet appear as stationary states, but only one corresponds to the true ground state, while the other is metastable. In a crossover, the transition is smooth, and near the transition, the ground state exhibits characteristics of both solitons and droplets~\cite{Pathak2022,cappellaro2018collective}.

So far, we have discussed energy competition in different atomic components. In dipolar quantum gases, a competition arises from two distinct sources of interaction: dipolar and short-range interactions. Owing to the interplay between them, a balance of forces can be achieved that stabilizes dipolar self-bound droplets~\cite{Baillie2017,bisset2021quantum,edmonds2020quantum}, analogous to Bose-Bose droplets discussed above. In recent years, these dipolar droplets have attracted significant attention, largely due to their close connection to the emergence of dipolar supersolids~\cite{Bottcher2019,tanzi2019supersolid,Chomaz2019,norcia2021two}. At the same time, dipolar bright solitons have been the subject of extensive theoretical studies~\cite{Gligori2008,Lakomy2012,Pedri2005a,Huang2017,Edmonds2017}.

The dipolar, single-component, long-lived system is an ideal platform to investigate transitions between solitons and droplets, while also offering a viable route to achieve quasi-2D solitons experimentally, a long-standing goal. An open question is whether the soliton–droplet transition and the associated bistability persist in quasi-2D, and whether this transition leaves identifiable dynamical signatures.

\begin{figure*}
    \centering
    \includegraphics[width=1\linewidth]{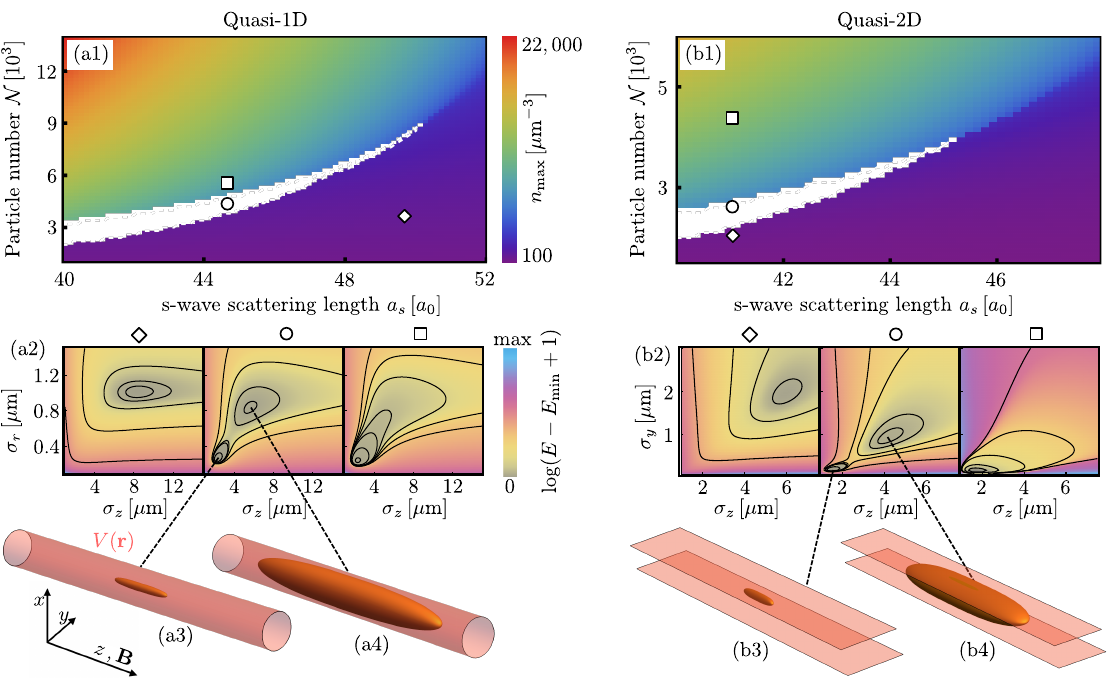}
    \caption{Soliton-to-droplet transition in a dipolar gas. Peak density of the VM ground state as a function of the s-wave scattering length $a_s$ and the particle number $\mathcal{N}$ for (a1) a quasi-1D and (b1) a quasi-2D geometry. White region denotes bistability. The trapping frequencies are $\omega_x=\omega_y=2\pi\times55\rm Hz$ in (a) and $\omega_x=2\pi\times180\rm Hz$ in (b). (a2),(b2) Energy surfaces in the $\sigma_{r,y}, \sigma_z$-plane, where $\sigma_r$ is the width in the transverse directions ($x,y$), for three points marked in the phase diagram showing a soliton ground state (diamond), a bistability (circle) and a droplet ground state (square). In the bistable region, the droplet (a3),(b3) and the soliton (a4),(b4) are both solution of the eGPE. The isosurfaces of the particle density (orange) are taken at values of half of the maximum density.}
    \label{fig:PhaseDia}
\end{figure*} 

To tackle these questions, we analyze a system of dipolar $^{166}\rm Er$ atoms confined in an infinite tube and an infinite plane potential. We demonstrate that the system can host both soliton and droplet states as its ground states. Furthermore, we report the presence of a crossover transition and a bistable region in the phase diagram, similar to the findings of Cheiney {\it et al.}~\cite{cheiney2018bright} from a Bose mixture. Our analysis is based on a variational model (VM), see Fig.\,\ref{fig:PhaseDia}, and is compared with numerical calculations of the ground state of the extended Gross-Pitaevskii equation (eGPE). We demonstrate that the transition including the bistability also exists in a quasi-2D trap, which is unique for the dipolar BEC. We show that the soliton-to-droplet transition is reflected in the structure factor, which can be further used experimentally to probe the transition. Figures~\ref{fig:PhaseDia}(a1),(b1) display the phase diagrams for the quasi-1D (left panels) and quasi-2D (right panels) setups, respectively, with color indicating the maximum density of the ground state. The white regions correspond to bistable regions, where there are two minima in the energy landscape of the system (see the middle panel in Fig.~\ref{fig:PhaseDia}(a2),(b2)). One minimum corresponds to a droplet (Fig.~\ref{fig:PhaseDia}(a3),(b3)), and the other to a soliton (Fig.~\ref{fig:PhaseDia}(a4),(b4)).\\
\indent In the following, we first introduce the theoretical framework employed in our paper, including the eGPE, the Bogoliubov–de Gennes (BdG) formalism, and the VM. We then present numerical results for an infinite tube (Sec.\,\ref{3}) and an infinite plane potential (Sec.\,\ref{4}), compare them with the predictions of the VM, and discuss the collective excitations of the system. After discussing the soliton–to-droplet transition by comparing our results with state-of-the-art experiments (Sec.\,\ref{5}), we conclude in Sec.\,\ref{6}.

\section{Theoretical Framework}
In this section we give an overview about the methods used in this work. We use a beyond mean-field description valid for weakly interacting Bosons at zero temperature. 

\subsection{Extended Gross-Pitaevskii equation}
The wave function $\Psi(\mathbf{r},t)$ of the system is described by the eGPE $i\hbar \partial \Psi/\partial t=\delta E[\Psi]/\delta \Psi^*\equiv \hat{H}\Psi$, where the energy functional is given by 
\begin{align}
\label{functional}
   \notag E=&\int \text{d}^3\mathbf{r} \Bigg(\frac{\hbar^2}{2m}|\nabla\Psi|^2+V|\Psi|^2+\frac{g}{2}|\Psi|^4\\&+\frac{1}{2}V_{\rm dd}(\mathbf{r})*|\Psi|^2+\frac{2\gamma_{\rm QF}}{5}|\Psi|^5\Bigg)\,\, .
\end{align}
Here, the contact interaction strength is given by $g=4\pi\hbar^2a_s/m$ with the s-wave scattering length $a_s$. The dipole-dipole potential takes the form $V_{\rm dd}(\mathbf{r})= g_{\rm dd}\frac{1-3\cos^2\theta}{|\mathbf{r}|^3}$ and its coefficient $g_{\rm dd}=3\hbar^2a_{\rm dd}/m$ is determined by the dipolar length $a_{\rm dd}=66.5 a_0$ for $^{166}\rm Er$~\cite{natale2022bloch}. The angle $\theta$ is given by the angle between $\mathbf{r}$ and the direction of the dipoles, here taken along $z$. We take into account beyond-mean-field corrections in the last term of Eq.~\eqref{functional}, where $\gamma_{\rm QF}=\frac{32}{3}g\sqrt{a_s^3/\pi}(1+\frac{3}{2}a_{\rm dd}^2/a_s^2)$. A quasi-1D setup can be realized with a harmonic confinement in the transverse directions, $V=m(\omega_x^2x^2+\omega_y^2y^2)/2$, whereas a quasi-2D setup corresponds to confinement in a single direction only, $V=m\omega_x^2x^2/2$. In this work, we fix the number of particles $\mathcal{N}$, which corresponds to a chemical potential $\mu=\int \text{d}^3\mathbf{r}\Psi_0^*\hat{H}\Psi_0/\mathcal{N}$. Note that in our work the 3D-LHY term can be used, because in our parameter regimes the system always fulfills $|\mu|> \hbar\omega_{x,y}$, which effectively makes the system three-dimensional~\cite{Petrov2000,Menotti2002}. The ground state $\Psi_0$ of the system can be computed by propagating an initial state $\Psi(\mathbf{r},0)$ in the eGPE in imaginary time. The calculation of the convolution term in Eq.~\eqref{functional} is carried out in Fourier space. To avoid Fourier aliasing we introduce a transverse cut-off, $\rho_c$, and a cut-off in $z$-direction, $Z_c$. By doing so we make use of the cylindrical shape of the wavefunction. A semi-analytical expression for the dipolar potential in momentum space is given in Appendix A.

\subsection{Bogoliubov - de Gennes equations}
The collective modes can be estimated by linearizing around the ground state $\Psi_0$ according to: $\Psi(\mathbf{r},t)=e^{-i\mu t/\hbar}(\Psi_0(\mathbf{r})+u_j(\mathbf{r})e^{-i\omega_jt}+v_j(\mathbf{r})^*e^{i\omega_jt})$. Any excitation is therefore defined by an energy $\hbar\omega_j$ and its amplitudes $(u_j,v_j)$. Inserting this term in the eGPE and retaining terms linear in $u_j,v_j$ yields the BdG equations~\cite{Pitaevskii16}. By introducing $f_j=u_j+v_j$ we can write them as a single equation
\begin{align}
\label{BdG}
    \left(\hat{H}-\mu\right)\left(\hat{H}-\mu+2\hat{X}\right)f_j=\hbar^2\omega_j^2f_j\,\, .
\end{align}
The operator $\hat{X}$ is defined by its action on a function $f$~\cite{Baillie2017},
\begin{align}
    \notag \hat{X}f(\mathbf{r})=&\Psi_0(\mathbf{r})\int \text{d}^3\mathbf{r}^\prime V_{\rm dd}(\mathbf{r}-\mathbf{r}^\prime)\Psi_0(\mathbf{r}^\prime)f(\mathbf{r^\prime})+g\Psi_0^2(\mathbf{r})f(\mathbf{r})\\&+\frac{3}{2}\gamma_{\rm QF}\Psi_0^3(\mathbf{r})f(\mathbf{r})\,\, .
\end{align}
Further, we have to consider the normalization of each mode arising from the Bose-Bose commutator relations and reading as 
\begin{align}
    \hbar\omega_j=\int d^3\mathbf{r}f_j\left(\hat{H}-\mu+2\hat{X}\right)f_j
\end{align}
for every $\omega_j\neq0$. We diagonalize Eq.~\eqref{BdG} using standard eigenvalue solvers to find the stable excitations of ground states obtained from the eGPE.
\subsection{Variational Model}\label{VM}
The ground state can also be calculated in a variational model by minimizing the energy in Eq.~\eqref{functional} directly, i.e. solving the equation $\delta E[\Psi]/\delta\Psi^*=0$. For this we substitute the ansatz $\Psi(\textbf{r})=\sqrt{\mathcal{N}}\phi(x,y)\psi(z)$ into Eq.~\eqref{functional}. The individual functions, normalized to unity, are assumed to be of the form

\begin{align}
\label{ansatz1}
\notag \phi(x,y)&=\sqrt{\frac{r_\rho}{2\pi\Gamma(2/r_\rho)\sigma_x\sigma_y}}e^{-\frac{1}{2}\left(x^2/\sigma_x^2+y^2/\sigma_y^2\right)^{r_\rho/2}}, \\
\psi(z)&=\sqrt{\frac{r_z}{2\Gamma(1/r_z)\sigma_z}}e^{-\frac{1}{2}\left(|z|/\sigma_z\right)^{r_z}}\, ,
\end{align}

with the Gamma function $\Gamma(x)$. The widths $\sigma_{j}$ and the exponents $r_\rho,r_z$ are variational parameters. This ansatz is a modified version of the one employed in Ref.~\cite{poli2021maintaining}, now allowing for radial asymmetry. When $r_\rho, r_z=2$, our ansatz is close to the analytically known shape of a soliton~\cite{Korteweg1895}. Droplets, on the other hand, are captured by higher exponents~\cite{Otajonov2019,Lavoine2021} resulting in a flat-top profile for high particle numbers. Substituting the ansatz from Eq.~\eqref{ansatz1} into the energy functional in Eq.~\eqref{functional} and integrating yields the total energy as a function of the variational parameters, see Appendix B. The ground state of the system is determined by the set of parameters $(\sigma_i,r_j)$ that minimize the sum of all single energy contributions and can be found with standard function minimizers.
\section{Quasi-1D Droplet to Soliton transition}\label{3}
The VM predicts a tricritical point located at
$(\mathcal{N}_T, a_{s,T}) \approx (9 \times 10^3, 50a_0)$, as shown in Fig.~\ref{fig:PhaseDia}(a1). At this point, the wave functions of the droplet and soliton merge, and the two states become indistinguishable. We expect bistability and hence a first-order phase transition for $a_s< a_{s,T}$ and a smooth crossover for $a_s > a_{s,T}$. 
\subsection{Crossover}
\begin{figure}[t]
    \centering
    \includegraphics[width=1\columnwidth]{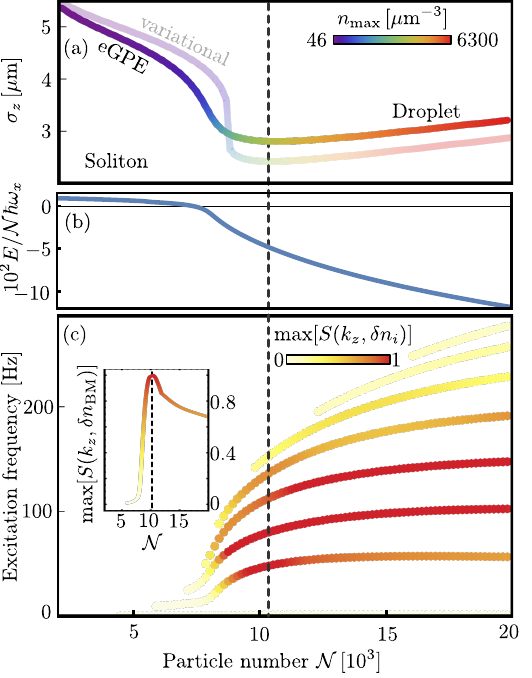}
    \caption{Soliton-to-droplet crossover in a quasi-1D system for $a_s=50a_0$ and $\omega_x=\omega_y=2\pi\times55\,\text{Hz}$. (a) Comparison between the width $\sigma_z$ of the ground state from the solution of the eGPE (solid line) and the one obtained from the VM (transparent line). The lines are colored with the peak density. (b) Energy of the ground state. (c) Excitations, where each mode frequency is colored by the maximum value of the structure factor (normed to unity). The inset shows the maximum value of the structure factor for the lowest-lying mode, the breathing mode.}
    \label{fig:1Dcross}
\end{figure}

Figure ~\ref{fig:1Dcross} shows our results for a fixed scattering length of $a_s = 50a_0$, obtained from the VM together with the solution of the eGPE. By calculating the width $\sigma_z$ of each ground state $\Psi_0$, we can distinguish between solitons and droplets, see Fig.~\ref{fig:1Dcross}(a). In droplets, the system approaches a maximum density and particles are added at the edges of the condensate. In solitons, on the other hand, adding particles increases the peak density. Consequently, for solitons we expect $\partial \sigma_z/\partial \mathcal{N}<0$ and for droplets $\partial\sigma_z/\partial\mathcal{N}>0$. The black dashed line marks the point where $\partial \sigma_z / \partial \mathcal{N} = 0$, located at $\mathcal{N} \approx 10^4$, as obtained from the solution of the eGPE.\\
\indent It is important to note that the transition between the droplet and soliton occurs smoothly. Near the crossover, the width of the soliton decreases rapidly, indicating proximity to the tricritical point. However, in the solution of the eGPE, the width decreases more gradually, suggesting that the critical $a_{s}$ for the crossover predicted by the VM is slightly overestimated. The color of the curves scales with the maximum density of the ground state. Close to the transition, the maximum density rises sharply, signaling the onset of mean-field collapse and droplet formation.\\
\indent Figure~\ref{fig:1Dcross}(b) shows the energy per particle, obtained from the eGPE. 
In the soliton regime, the energy is positive and varies weakly with $\mathcal{N}$. In the limit of small particle numbers, the total energy approaches $\mathcal{N\hbar\omega}$, as for $\mathcal{N}$ uncoupled two-dimensional isotropic harmonic oscillators. In contrast, in the droplet regime, the energy decreases rapidly, indicating that the interaction energy dominates the total energy of the system. In the case of a one-dimensional trap, we can derive a stability condition for self-bound solutions from the variational model. There, the kinetic energy scales with $1/\sigma_z^2$ and the interaction energy is dominated by terms of the order $1/\sigma_z$. This energy allows for a bound solution only if the interactions are overall attractive, i.e. $a_{\rm dd}>a_s$. If this condition is fulfilled, we expect bound solutions to appear for all particle numbers. On the other hand, the particle number of the transition varies with $a_s$. We can understand this by simple arguments. At the transition, the mean field interaction energy equals the kinetic energy. Since the mean field energy scales with $(a_{\rm dd}-a_s)\mathcal{N}^2$ and the kinetic energy with $\mathcal{N}$, the particle number of the transition shifts towards larger values when the s-wave scattering length is increased, see Fig.\,\ref{fig:PhaseDia}. With the same arguments, we can also estimate the effect of increasing the trapping frequency. Eventually, tightening the trap leads to a higher peak density and a higher interaction energy. Consequently, the transition happens at lower particle numbers, which is also confirmed by the VM. \\
\indent Finally, Fig.~\ref{fig:1Dcross}(c) shows the excitation spectrum of the system as a function of $\mathcal{N}$. We display only stable modes, i.e., those satisfying the boundary condition $f \rightarrow 0$ for $z \rightarrow \pm \infty$. As the number of particles increases, a larger number of stable modes appear. A threshold energy marks the onset of a continuous spectrum. This threshold corresponds to the self-binding energy per particle, above which particles escape the condensate. The self-binding energy differs from the chemical potential as a result of the trapping potential and the quasi-low-dimensional geometry.\\
\indent Each excitation mode is color-coded according to the maximum value of its structure factor, $
S(k_z, \delta n_i) = \left|\mathcal{F}_z\{\delta n_i(\mathbf{r})\}\right|^2$, where $\delta n_i = \Psi_0 f_i$ represents the density fluctuations, and $\mathcal{F}_z$ denotes the Fourier transformation in $z$ direction. 
All modes shown cause dynamics along the dipolar direction ($z$). Their spatial density profiles are discussed in Appendix C. The inset displays the maximum of the structure factor for the lowest-lying mode, the breathing mode, which exhibits a pronounced maximum at the crossover. Since the breathing mode affects the axial width $\sigma_z$, this mode is directly related to the soliton-to–droplet transition. The maximum response of the breathing mode provides a directly accessible experimental signature of the soliton–to-droplet transition. Elementary excitations can be initially excited either via Bragg spectroscopy~\cite{Stenger1999,blakie2002theory,Houwman2024} or through modulation of the interaction strength~\cite{Pollack2010,Haller2009} or the trapping frequencies~\cite{DiCarli2019}. The modes above the breathing mode in Fig.~\ref{fig:1Dcross} are higher-lying multipole modes. In addition to those, we also found the dipole mode arising from the confinement in $x$ and two Goldstone modes: one associated with the spontaneously broken $U(1)$ symmetry and another corresponding to the spontaneously broken translational symmetry along the $z$-axis. The latter mode becomes gapped if we add a confinement in the $z$-direction~\cite{guo2019low}. 
\subsection{Bistability}
\begin{figure}[t]
    \centering
    \includegraphics[width=1\columnwidth]{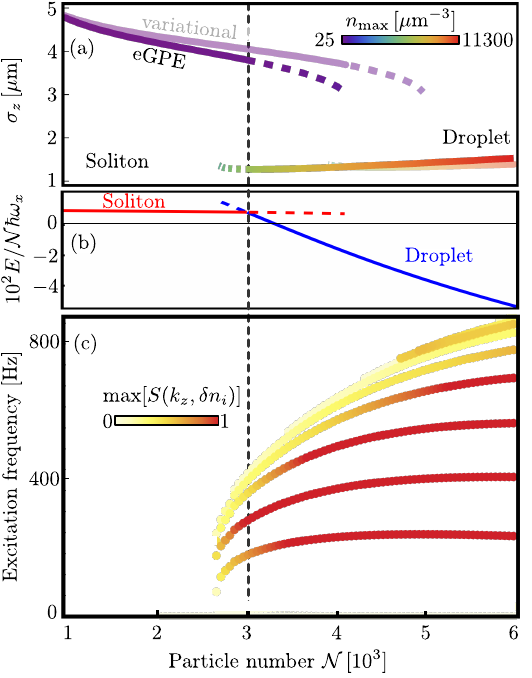}
    \caption{First-order phase transition in a quasi-1D system between a soliton and a droplet for $a_s=45a_0$ and $\omega_x=\omega_y=2\pi\times55\,\text{Hz}$. (a) The width $\sigma_z$ calculated from the solution of the eGPE (solid line) and the VM (transparent line). Metastable states are indicated by the dashed line. The color of the curves corresponds to the peak density. (b) Energy as a function of $\mathcal{N}$. (c) Excitation spectrum, where each mode frequency is colored according to its contribution to the structure factor.}
    \label{fig:1Dbi}
\end{figure}

For values of $a_s$ smaller than $a_{s,T}$, the VM predicts a bistable region, i.e. soliton and droplet are both stationary solutions of the system. We investigate this region of the phase diagram by setting the scattering length to $a_s=45a_0$. We show the width as a function of the particle number in Fig.~\ref{fig:1Dbi}(a). Again, the ground state is shown as a solid line, while we indicate metastable states by dashed lines. Metastable states can be found in the imaginary time propagation by selecting initial conditions that are already close to these states. Practically, we have propagated an initial state in imaginary time until convergence is reached and then we have increased/decreased the particle number carefully so as not to push the soliton/droplet out of the local energy minimum. Indeed, the eGPE confirms the bistability of the two states. The true ground state can then only be found by comparing the energies, Fig.~\ref{fig:1Dbi}(b). In fact, the phase transition is found to be of first-order. It is straightforward to determine the corresponding particle number, $\mathcal{N}\approx3\times 10^3$ in this case. According to the VM, the phase transition occurs at \(\mathcal{N} \approx 4.3 \times 10^3\), a value notably higher than the one from the solution of the eGPE, consistent with previous observations for VMs~\cite{Baillie2017}.\\
\indent The presence of a phase transition makes it possible to clearly distinguish between solitons and droplets. In the droplet phase itself, the system may exhibit two distinct phases of self-binding. For small particle numbers the droplet is self-bound in the $z$-direction only. This especially holds for the metastable droplet states, where the total energy exceeds the trapping energy. The potential is essential for the existence of these states, as it balances their kinetic energy. For large particle numbers, however, the energy becomes negative, and the trap is, in principle, not required to stabilize the system~\cite{ferrier2018scissors}.\\
\indent We present the collective excitations in Fig.~\ref{fig:1Dbi}(c). Here, we only show the excitations that correspond to the droplet. Consequently, for $\mathcal{N}<3\times 10^3$, we have linearized around a state, which is not a ground state. However, all modes remain real, which implies that we can only quantify the instability of a metastable state by considering nonlinear effects. For all soliton solutions, there exists no stable excitation. The reason for this is the relatively flat energy surface of the soliton, $\partial^2 E/\partial\sigma^2_i|_{\rm droplet}\gg\partial^2 E/\partial\sigma^2_i|_{\rm soliton}$, also visible in  Fig.~\ref{fig:PhaseDia}(a2),(b2). The presence of stable excitations thus offers a clear criterion to distinguish droplets from solitons~\cite{Malkin1991,bakkali2023cross,cappellaro2018collective}. 
\section{Quasi-2D Droplet to Soliton transition}\label{4}
Following the procedure from the previous section, we examine the formation of droplets and solitons in a quasi-2D trap, and reveal that also in this case there exists a bistable and a crossover region. For this we fix the trapping frequency to $\omega_x=180$Hz. Before exploring the crossover, we discuss the stability of 2D solitons.\\
\begin{figure}[b]
    \centering
    \includegraphics[width=1\columnwidth]{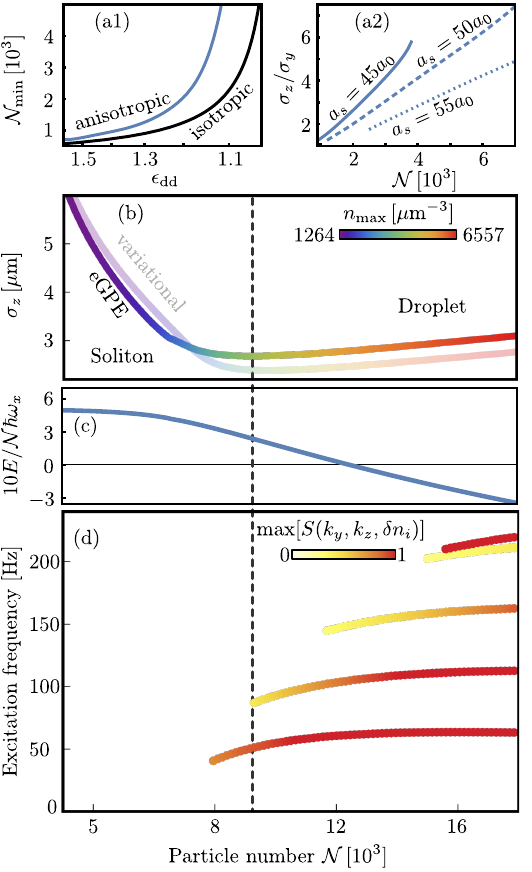}
    \caption{(a1) Critical atom number for 2D-solitons as a function of $\epsilon_{\rm dd}$ in a cylindrical symmetric system (black line) and for an anisotropic soliton (blue line) obtained from the VM. We define $\epsilon_{\rm dd}=a_{\rm dd}/a_s$ for the anisotropic/dipolar case and $\epsilon_{\rm dd}=a_{\rm dd}/2 a_s$ for the isotropic/anti-dipolar case. (a2) Aspect ratio $\sigma_z/\sigma_y$ as a function of the particle number $\mathcal{N}$ for different values of the $a_s$ for the anisotropic soliton obtained from the VM. (b) Width $\sigma_z$ of the ground state as a function of $\mathcal{N}$ for the solution of the eGPE (solid line) and the VM (transparent line) for $a=50a_0$ and $\omega_x=2\pi\times180\,\text{Hz}$. The color indicates the value of the peak density. (c) Energy as a function of $\mathcal{N}$. (d) Excitation frequencies as a function of $\mathcal{N}$ with their contribution to the structure factor indicated by color.}
    \label{2DCross}
\end{figure}
\indent In the quasi-2D set-up, the system is confined in one direction only. Due to the absence of attractive interactions in the $y$-direction, the stability of the system is weakened. This becomes evident by looking at the total energy of the system using the VM. We assume that there exist solutions of large width $\sigma_y,\sigma_z\sim\sigma\gg\sigma_x$. Then, the kinetic energy and the short-range interaction scale with $1/\sigma^2$. The dipolar energy scales with $-1/\sigma^2$ in the $0^{\rm{th}}$ order approximation of $f_{\rm dip}$, defined in Appendix B. To obtain bound solutions we have to take into account higher orders of $f_{\rm dip}$. The next order term of the dipolar energy goes with $b/\sigma^3$, where $b$ is a positive constant. Taking it all together, the total energy reads as $E\sim a/\sigma^2+b/\sigma^3$, which allows for bound solutions only when $a<0$. Practically, this condition can be written in the form

\begin{align}
\label{Inequality}
\frac{E_{\text{kin},y,z}}{E_{\rm sr}} < \frac{a_{\rm dd}}{a_s} - 1 \,.
\end{align}

where $E_{\text{kin},y,z}$ and $E_{\rm sr}$ are the kinetic energy associated with the $y$ and $z$-directions, and the energy contribution arising from short-range interactions, respectively. Note that the left-hand side of Eq.~\eqref{Inequality} depends explicitly on $\sigma_y$ and $\sigma_z$. The equation implies that there exists not only an upper bound for the s-wave scattering length ($a_s<a_{\rm dd}$) but also a minimum atom number $\mathcal{N}_{\rm min}$ for 2D dipolar solitons, which is fundamentally different from the quasi-1D behavior. We can find a lower bound on the critical particle number by assuming $\sigma_y=\sigma_z$ in Eq.~\eqref{Inequality}. Then, we obtain $\mathcal{N}_{\rm min}=\sqrt{2}\sigma_x/\sqrt{\pi}(a_{\rm dd}-a_s)$. Interestingly, this result becomes equivalent to Eq.~(6) in Ref.~\cite{Pedri2005a} when replacing $a_{\rm dd} \rightarrow a_{\rm dd}/2$. In this reference, the authors studied dipolar bright solitons in a 2D trap, where the magnetic field $\mathbf{B}$ is oriented parallel to the confinement direction $x$ and rotated rapidly, resulting in isotropic solitons in an anti-dipolar condensate. The factor $1/2$ arises because anti-dipolar interactions are effectively weaker than dipolar interactions. Anisotropic solitons were studied in Ref.~\cite{tikhonenkov2008}; the authors derived a condition for the interaction strength. However, stability with respect to the particle number was not discussed.\\
\indent In Fig.~\ref{2DCross}(a1), we compare the minimum atom number $\mathcal{N}_{\rm min}$ obtained from our VM for an anisotropic soliton, where $\mathbf{B}=B\mathbf{e}_z$, to that of an isotropic soliton where the magnetic field is aligned along $x$ and is rotated with high frequency. Solitons are stable for $\mathcal{N}>\mathcal{N}_{\rm min}$. Close to the instability, significant differences in $\mathcal{N}_{\rm min}$ arise between the anisotropic and isotropic solitons, indicating that the stability of the anisotropic soliton is limited to a relatively smaller parameter region. We also show the ratio $\sigma_z/\sigma_y$, see Fig.~\ref{2DCross}(a2) . Indeed, in our set-up, there exist three different length scales. At the soliton-to-droplet transition, $\sigma_y$ approaches $\sigma_x$, resulting in an increasing ratio $\sigma_z/\sigma_y$ towards higher particle numbers. Consequently, there exists no anisotropic droplet with $\sigma_x\neq\sigma_y$ in our study. We found that the anisotropy of the 2D soliton is maximized for $(\mathcal{N},a_s)\approx (7\times10^3,50a_0)$, where it has a cigar-like shape. \\
\subsection{Crossover}
\indent To investigate the crossover, we set $a_s=50a_0$ and choose $\mathbf{B}=B\mathbf{e}_z$. Fig.~\ref{2DCross}(b) shows the width as a function of $\mathcal{N}$, where we observe a nice agreement between the VM and the solution of the eGPE. The energy is presented in Fig.~\ref{2DCross}(c) and approaches $\mathcal{N}\hbar\omega_x/2$. In contrast to the quasi-1D case, the soliton energy exceeds this value for $\mathcal{N}<\mathcal{N}_c$, which means that the system is not self-bound below the critical atom number. As a consequence of the anisotropy of the system, there are two spontaneously broken translational symmetries, and we observe, in total, three Goldstone modes. Figure~\ref{2DCross}(d) shows all stable excitations. Again, the lowest-lying mode is the breathing mode of the dipolar direction. The fifth mode is the quadrupole mode and is particularly pronounced in the structure factor, see Appendix C for further details. In the present case, the quadrupole mode has a higher energy compared to the lowest lying breathing mode. This reflects the well-known fact that droplets of low particle numbers are highly compressible~\cite{Stringari1996}.\\
\indent Here, we do not observe any clear signature of the transition in the system’s structure factor. The maximum value of the structure factor increases monotonically with $\mathcal{N}$ for all modes. This behavior can, however, be attributed to the anisotropy of the system. The breathing mode predominantly affects the width along the dipolar direction. At the transition, not only does the width of the dipolar direction, $\sigma_z$, change, but the width along the $y$-direction, $\sigma_y$, also varies. Consequently, the breathing mode in two dimensions cannot be directly associated with the transition between the two states.\\
\subsection{Bistability}
\begin{figure}[t]
    \centering
    \includegraphics[width=1\columnwidth]{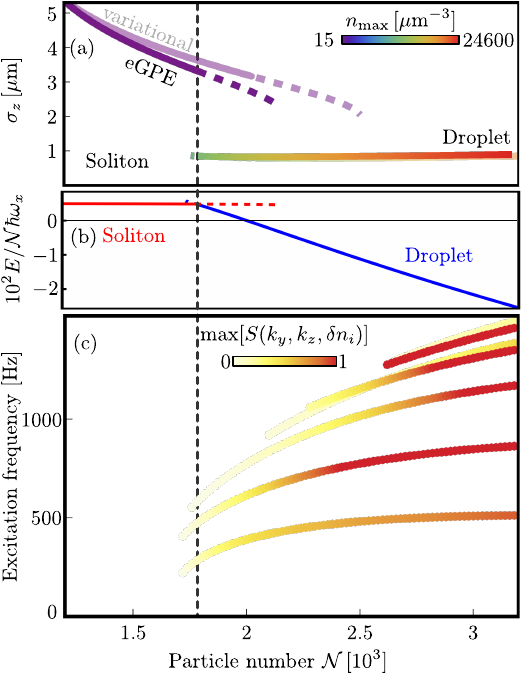}
    \caption{First-order phase transition from a soliton to a droplet in an infinite plane potential. (a) Width $\sigma_z$ of the ground state as a function of $\mathcal{N}$, results from the solution of the eGPE and the VM are shown as a solid and transparent line, respectively. The color of each curve is given by the value of the peak density of the state. The dashed lines correspond to metastable states. (b) Energy as a function of $\mathcal{N}$. (c) Excitation frequencies, each colored by the maximum value of the structure factor.}
    \label{2DBistable}
\end{figure}

We fix $a_s=40a_0$ to investigate the bistability region further. For these parameters, the eGPE predicts the phase transition at $\mathcal{N}\approx 1800$, whereas it occurs at $\mathcal{N}\approx 2100$ in the VM, see Fig.~\ref{2DBistable}(a). The droplet is metastable only very close to the transition, while we observe a metastable soliton up to $\mathcal{N} \approx 2200$. Notably, the ground states calculated from the VM in the droplet phase are very close to the solutions of the eGPW. However, the VM finds metastable solitons up to $\mathcal{N}\approx2500$. From the energy in Fig.~\ref{2DBistable}(b), we can identify the particle number at the phase transition. In the excitation spectrum in Fig.~\ref{2DBistable}(c), we find the breathing mode and the quadrupole mode, first and fourth mode respectively, see also Appendix C. As in the 1D case, there exists no stable mode in the soliton regime, and we show the modes of the metastable droplet for $\mathcal{N}$ below the transition.  

\section{Experimental Realization}\label{5}
So far, we have studied solitons and droplets in the absence of trapping in either one direction (quasi-1D) or two (quasi-2D). In experiments, however, it is often necessary to confine the system in all spatial directions. Here, we aim to investigate the impact of a three-dimensional trap on the transition from a droplet to a soliton in a cigar shaped trap. Our findings also provide new insight into the experimental observations reported by Chomaz \textit{et al.}~\cite{chomaz2016quantum}. In their work, the authors measured the number of atoms in a dipolar condensate as a function of time for the regime $g_{\rm dd} > g$, see Fig.~4(c) in their paper. Initially, the system contained $10^5$ atoms, but this number gradually decreased due to losses. The rapid decay at early times can be attributed to the high density of the initial droplet. However, the authors observed a subsequent slowing of the decay, leading to a long-lived state with approximately $5 \times 10^3$ atoms and significantly reduced losses.\\
\indent Our zero-temperature simulations indicate that, despite the presence of a three-dimensional trap, the system can indeed support solitons under the experimental parameters reported in their study. However, the existence of solitons is limited by the weak trap in the dipolar direction. For low particle numbers and when $a_s\rightarrow a_{\rm dd}$, the soliton width increases, leading to a rise in potential energy. Consequently, it is energetically favorable to transition to the normal BEC phase. The presence of a soliton is in perfect agreement with the observation of long-lived states for an intermediate range of particle numbers and when $a_s$ is significantly smaller than $a_{\rm dd}$, see their Fig~4(c). Thus, the behavior observed by Chomaz \textit{et al.}~\cite{chomaz2016quantum} likely corresponds to the transition from a droplet to a soliton.\\
\indent Because of their reduced stability, producing two-dimensional solitons remains experimentally challenging. In non-dipolar condensates, two-dimensional Townes solitons have already been observed~\cite{Chen2021,Bakkali2021}. To the best of the authors’ knowledge, however, 2D dipolar solitons have not yet been observed. The results reported in Ref.~\cite{natale2022bloch} suggest that a soliton-like state may have been realized in a quasi-2D trap, but additional experiments are needed to confirm whether the observed state is indeed a genuine soliton. It is worth mentioning that the bistable region may also be observed in other dipolar condensates. In Ref.~\cite{ferrier2018scissors}, the authors comment on the soliton-to-droplet transition in quasi-2D condensate of $^{164}\rm Dy$ atoms. The sharp transition at low particle numbers, see their Fig.~S1, likely implies a first-order phase transition connected to a bistability of soliton and droplet.
\section{Conclusion}\label{6}
In this work, we have investigated the transition from solitons to droplets in a quasi-1D and quasi-2D trap. Our results reveal the presence of a bistable region where both states can coexist, as well as a smooth crossover between them, depending on the particle number and interaction strength. We employed both the VM and imaginary time propagation to study these transitions. While the VM provides accurate results deep within the soliton and droplet regimes, it becomes less reliable near the transition region. Nevertheless, it successfully predicts the existence of the bistable region and offers a good approximation of the ground state.\\
\indent Our study bridges the gap between the findings of Ref.~\cite{cheiney2018bright}, which explored transitions in non-dipolar Bose–Bose mixtures, and the extensive prior research on dipolar bright solitons and droplets~\cite{Bttcher2021}. In quasi-1D, we report a maximum response of the structure factor for the breathing mode at the transition, which can be used experimentally to probe the transition. For the system considered here, we found that bistability can occur not only in a quasi-1D trap but also in a quasi-2D geometry. Furthermore, we analyzed the system’s excitations: while a breathing mode appears in the quasi-1D case, an additional quadrupole mode emerges in quasi-2D configurations. Finally, we discussed the stability of 2D dipolar bright solitons and proposed possible experimental platforms for observing bistability and realizing 2D dipolar bright solitons. In a 1D-system, the transition may have already been observed as our comparison to the experiments in Ref.~\cite{chomaz2016quantum} show. 

\section{Acknowledgments}
We are grateful to Alfonso Annarelli and Francesco Formicola for their early contributions to this work through the AURORA excellence fellowship program. 
This work is supported from the Knut and Alice Wallenberg Foundation (Grants No. KAW2018.0217 and No. KAW2023.0322) and the Swedish Research Council (Grant No. 2022-03654). The Innsbruck team acknowledges support from the European Research Council through the Advanced Grant DyMETEr (10.3030/101054500), a NextGeneration EU Grant AQuSIM through the Austrian Research Promotion Agency (FFG) (No.\,FO999896041), and the Austrian Science Fund (FWF) Grant (10.55776/PAT1597224) and the Cluster of Excellence QuantA (10.55776/COE1).

\section*{Appendix A: Numerical Details for the solutions of the eGPE}\label{A1}
It is most convenient to calculate the dipolar part of the time evolution operator in Fourier space. However, we need to be careful with Fourier imaging. Here, we use a cylindrical cut-off reflecting the cigar/pancake shape of the ground state. By doing so we obtain the semi-analytic expression~\cite{Lu2010}
\begin{align}
    \notag &\tilde{V}_{\rm dd}^{\rho_c,Z_c} (\mathbf{k})=4\pi g_{\rm dd}\Big(\cos^2\theta_k-\frac{1}{3}\\ \notag &+e^{-Z_ck_\rho}\left[\sin^2\theta_k\cos(Z_c k_z)-\sin \theta_k \cos \theta_k \sin(Z_ck_z)\right]\\ &-\int_{\rho_c}^\infty\rho  \text{d}\rho\int_0^{Z_c}\text{d} z\cos (k_z z)\frac{\rho^2-2z^2}{(\rho^2+z^2)^{5/2}}J_0(k_\rho\rho)\Big)\,\, ,
\end{align}
where we denote the first Bessel function of the first kind by $J_0$ and $\rho=\sqrt{x^2+y^2}$. The potential only depends on the parameters of the numerical grid, and it must be calculated once before starting the propagation. Here, we use $(\rho_c,Z_c)=(L_x/2,L_z/2)$ for a box with $L_x=L_y$. For the anisotropic dipolar 2D soliton, we must be careful, as it must be $\Psi=0$ for $|x|,|y|>L_x/4$ and $|z|>L_z/4$. We propagate in imaginary time via a split-time Fourier spectral algorithm~\cite{Bao2003} on a grid with the dimensions $96\times 96\times256$.
\section*{Appendix B: Calculation of the energy in the VM}
We can find the total energy
\begin{align}
    E = E_\text{kin} + E_\text{trap} + E_\text{sr} + E_\text{dd} + E_\text{qf}\, \label{Eq:EsdAppend}
\end{align}
by integrating out the single energy contributions
\begin{align}
\begin{aligned}
    E_\text{kin} &= -\frac{\hbar^2}{2m}\int \rm \text{d}^3 \mathbf{r}\, \Psi^*\nabla^2\Psi\,,\\
    E_\text{trap} &= \frac{m}{2}\int \text{d}^3 \mathbf{r}\, |\Psi|^2\sum_i\omega_i^2x_i^2\,,\\
    E_\text{sr} &= \frac{g}{2}\int \text{d}^3 \mathbf{r}\, |\Psi|^4\,, \\
    E_\text{dd} &= \frac{g_{\rm dd}}{2}\int \frac{\text{d}^3\textbf{k}}{(2\pi)^3} \left ( \frac{3k_z^2}{k^2} -1 \right ) |\tilde n(\textbf{k})|^2\,, \\
    E_\text{qf} &= \frac25\gamma_{\text{QF}}\int \rm d \mathbf{r}^3\,|\Psi|^5\,\, ,
\end{aligned}
\label{eqn:energies}
\end{align}
 corresponding to the kinetic, trap, short-range interaction, dipole-dipole interaction, and quantum fluctuation contributions, respectively. The density in Fourier space can be calculated via $\tilde n(\textbf{k})=\int \rm d \mathbf{r}^3\, e^{-i\textbf{k}\cdot\textbf{r}} |\Psi(\textbf{r})|^2$.\\
\indent After inserting Eq.~\eqref{ansatz1}, the kinetic energy reads as
\begin{align}
\frac{E_\text{kin}}{\mathcal{N}}=\frac{\hbar^2}{8m}\left(\frac{\sigma_x^2+\sigma_y^2}{\sigma_x^2 \sigma_y^2}\frac{r_\rho^2}{2\Gamma(2/r_\rho)} + \frac{r_z^2 \Gamma(2-1/r_z)}{\sigma_z^2 \Gamma(1/r_z)}\right)\,.
\end{align}
We obtain 
\begin{align}
\frac{E_\text{trap}}{\mathcal{N}}=\frac{m}{2}\left[\frac{\Gamma(4/r_\rho)}{2\Gamma(2/r_\rho)}\left(\omega_x^2\sigma_x^2 + \omega_y^2\sigma_y^2\right) + \frac{\Gamma(3/r_z)}{\Gamma(1/r_z)}\omega_z^2\sigma_z^2\right]
\end{align}
for the trap energy and the contribution from short range interactions is given by
\begin{align}
\frac{E_\text{sr}}{\mathcal{N}} = \frac{gn_0}{2^{1+2/r_\rho + 1/r_z}}\,\, .
\end{align}
Here, the central density is defined by
\begin{align}
    n_0 = \frac{\mathcal{N}r_\rho r_z}{4\pi \sigma_x\sigma_y\sigma_z\Gamma(2/r_\rho)\Gamma(1/r_z)}\,\, .
\end{align}
The energy contribution for the quantum fluctuations is
\begin{align}
\frac{E_\text{qf}}{\mathcal{N}} = \left(\frac25\right)^{1 + 2/r_\rho + 1/r_z}\gamma_\text{QF}n_0^{3/2}\,\, .
\end{align}
Due to self interaction in mean field theories, the integral for the dipolar energy can only be computed in the Fourier space. For this, we decompose the density in Fourier space via $\tilde n(\textbf{k})=\tilde n_\rho(k_x,k_y)\tilde n_z(k_z)$, with
\begin{align}
\begin{aligned}
\tilde n_\rho(k_x,k_y)&=\frac{r_\rho}{\Gamma(2/r_\rho)}\int_0^\infty \text{d}\rho\, \rho e^{-\rho^{r_\rho}} J_0\left(\sqrt{k_x^2\sigma_x^2+k_y^2\sigma_y^2}\rho\right)\,, \\
\tilde n_z(k_z) &= \frac{r_z}{\Gamma(1/r_z)} \int_0^\infty \text{d}z\, e^{-z^{r_z}}\cos(k_z\sigma_z z)\,\, .
\end{aligned}
\label{eqn:nk}
\end{align}
For the parameter regime used in this paper, we can approximate the Fourier transform of the density as~\cite{poli2021maintaining}
\begin{align}
\label{approxi}
 \tilde n(\mathbf{k})\simeq e^{-\alpha_\rho (k_x\sigma_x)^2-\alpha_\rho(k_y\sigma_y)^2 -\alpha_z(k_z\sigma_z)^2}   
\end{align}
where the functions $\alpha_\rho$ and $\alpha_z$ are functions that explicitly depend on the exponents $r_\rho,r_z$. They can be found by fitting Eqs.~\eqref{approxi} to Eqs.~\eqref{eqn:nk}. We note that the approximation made in Eqs.~\eqref{approxi} is well justified for the transition region, when $r_\rho,r_z\approx 2$. For large particle numbers, however, especially when the particle density has a flat top shape, the approximation breaks down. All our calculations are performed far away from this regime, and the good agreement between the VM and the exact solutions of the eGPE particularly in the droplet regime justifies the use of Eqs.~\eqref{approxi}. We can then calculate the dipolar energy and obtain
\begin{figure}[b!]
    \centering
    \includegraphics[width=1\columnwidth]{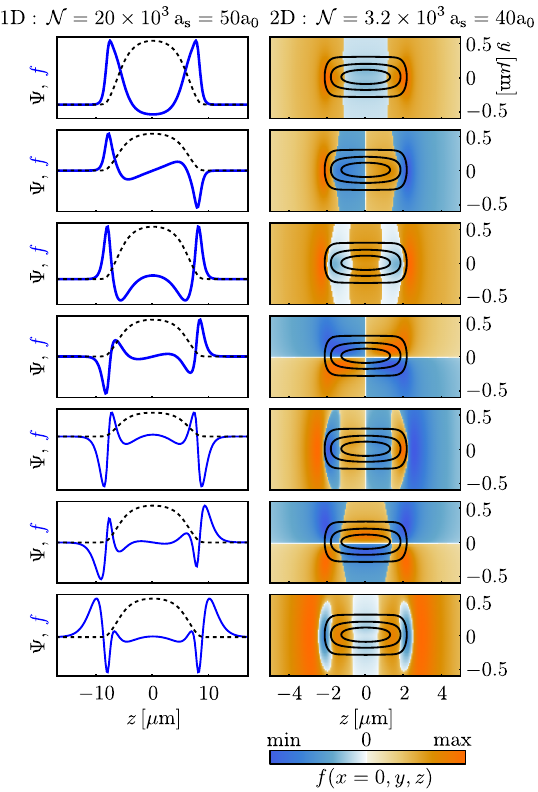}
    \caption{Stable modes calculated by solving the BdG-equations. Left Column: Quasi-1D. The condensate density is shown as a black dashed line and the mode function along $z$ in blue. For better visualization, we have normed both functions to the same amplitude. Right Column: Quasi-2D modes in the $yz$-plane. The value of the mode function corresponds to the color. We show the isolines of the density as black lines.}
    \label{AllModes}
\end{figure}
\begin{align}
\begin{aligned}
\frac{E_\text{dd}}{\mathcal{N}}=\frac{g a_\text{dd} \mathcal{N} }{2(2\pi)^{3/2}\ell_x\ell_y\ell_z a_s} f_\text{dip}\left(\frac{\ell_x}{\ell_z},\frac{\ell_y}{\ell_z}\right)\,,
\end{aligned}
\end{align}
where we have defined $(\ell_x^2,\ell_y^2,\ell_z^2)=4(\alpha_\rho\sigma_x^2,\alpha_\rho\sigma_y^2,\alpha_z\sigma_z^2)$. The function 
\begin{align}
\notag f_\text{dip}(\eta,\kappa) = & -1+\frac{3\kappa\eta}{\sqrt{1-\kappa^2}(1-\eta^2)}\times\\\ & \notag \Bigg[ F\left(\text{arcsin}\sqrt{1-\kappa^2}\Bigg|\frac{1-\eta^2}{1-\kappa^2}\right)\\ & \notag -E\left(\text{arcsin}\sqrt{1-\kappa^2}\Bigg|\frac{1-\eta^2}{1-\kappa^2}\right)\Bigg]\,,
\end{align}
was introduced in ~\cite{giovanazzi2006expansion,Glaum2007A} and makes use of the incomplete elliptic function of the first and second kind, $F(\phi|m)$ and $E(\phi|m)$ respectively. 

\section*{Appendix C: Mode functions of the collective excitations}

Solving the BdG equations yields the mode functions 
$f_i$, which are related to the density fluctuations through $\delta n_i(\mathbf{r},t)\sim \Psi_0f_i\cos(\omega_i t)$. In Fig.~\ref{AllModes}, left column, we present the spatial profiles of the modes, together with the ground state, for the quasi-1D setup, ordered by their energy from top to bottom. The first mode corresponds to the monopole (breathing) mode. The emergence of a hierarchical structure of even and odd collective modes is well known for quantum droplets~\cite{Blakie2023,Tylutki2020}. The corresponding modes of the quasi-2D droplet are shown in the right column. The first three modes show dynamics exclusively along the dipolar direction, of which the breathing mode is the lowest in energy. In addition, the quadrupole mode appears (fourth mode from the top), characterized by its angular dependence $f\sim \cos(2\phi)$. Interestingly, we also identify a higher-lying quadrupole mode, which can be constructed from the product of the second breathing mode and the first quadrupole mode (sixth mode from the top).
\newpage
\bibliography{refs1}

@article{Sivaram2021,
  title = {Is the binding energy of galaxies related to their core black hole mass?},
  author = {C. Sivaram, K. Arun},
  journal = {Journal of Astrophysics and Astronomy},
  volume = {42},
  issue = {2},
  year = {2021},
  url = {https://doi.org/10.1007/s12036-021-09732-4}
}

@article{Bulgac2002,
  title = {Dilute Quantum Droplets},
  author = {Bulgac, Aurel},
  journal = {Phys. Rev. Lett.},
  volume = {89},
  issue = {5},
  pages = {050402},
  numpages = {4},
  year = {2002},
  month = {Jul},
  publisher = {American Physical Society},
  doi = {10.1103/PhysRevLett.89.050402},
  url = {https://link.aps.org/doi/10.1103/PhysRevLett.89.050402}
}

@book{Tielens2005, place={Cambridge}, title={The Physics and Chemistry of the Interstellar Medium}, publisher={Cambridge University Press}, author={Tielens, A. G. G. M.}, year={2005}}

@article{Young1805,
  title = {An essay on the cohesion of fluids},
  author = {T. Young},
  journal = {Phil. Trans. R. Soc.},
  volume = {95},
  pages = {65-87},
  year = {2021},
  url = {https://royalsocietypublishing.org/doi/10.1098/rstl.1805.0005}
}

@article{Sutton1988,
doi = {10.1088/0022-3719/21/1/007},
url = {https://dx.doi.org/10.1088/0022-3719/21/1/007},
year = {1988},
month = {jan},
publisher = {},
volume = {21},
number = {1},
pages = {35},
author = {A P Sutton and M W Finnis and D G Pettifor and Y Ohta},
title = {The tight-binding bond model},
journal = {Journal of Physics C: Solid State Physics}
}

@article{gamow1930,
  title={Mass defect curve and nuclear constitution},
  author={G. Gamow},
  journal={Proc. R. Soc. Lond. A},
  volume={126},
  pages={632--644},
  year={1930}
}

@article{Duree1993,
  title = {Observation of self-trapping of an optical beam due to the photorefractive effect},
  author = {Duree, Galen C. and Shultz, John L. and Salamo, Gregory J. and Segev, Mordechai and Yariv, Amnon and Crosignani, Bruno and Di Porto, Paolo and Sharp, Edward J. and Neurgaonkar, Ratnakar R.},
  journal = {Phys. Rev. Lett.},
  volume = {71},
  issue = {4},
  pages = {533--536},
  numpages = {0},
  year = {1993},
  month = {Jul},
  publisher = {American Physical Society},
  doi = {10.1103/PhysRevLett.71.533},
  url = {https://link.aps.org/doi/10.1103/PhysRevLett.71.533}
}

@book{Drazin1989, place={Cambridge}, edition={2}, series={Cambridge Texts in Applied Mathematics}, title={Solitons: An Introduction}, publisher={Cambridge University Press}, author={Drazin, P. G. and Johnson, R. S.}, year={1989}, collection={Cambridge Texts in Applied Mathematics}}

@article{Zakharov1973,
       author = {Zakharov, V.~E. and Shabat, A.~B.},
        title = {Interaction between solitons in a stable medium},
      journal = {Soviet Journal of Experimental and Theoretical Physics},
         year = 1973,
        month = nov,
       volume = {37},
        pages = {823},
       adsurl = {https://ui.adsabs.harvard.edu/abs/1973JETP...37..823Z},
       url = {https://www.jetp.ras.ru/cgi-bin/dn/e_037_05_0823.pdf}
}

@article{Zakharov1979,
  title = {Integration of nonlinear equations of mathematical physics by the method of inverse scattering},
  author = {V. E. Zakharov and A. B. Shabat },
  journal = {Funct. Anal. Appl.},
  volume = {13},
  issue = {3},
  pages = {13--22},
  year = {1979},
  publisher = {},
  url = {https://link.springer.com/article/10.1007/BF01077483}
}

@book{Rajaraman1982,
  author    = {Rajaraman, R.},
  title     = {Solitons and Instantons: An Introduction to Solitons and Instantons in Quantum Field Theory},
  publisher = {North-Holland Publishing Company},
  address   = {Amsterdam, Netherlands},
  year      = {1982},
  isbn      = {0444862293}
}

@article{Emplit1987,
title = {Picosecond steps and dark pulses through nonlinear single mode fibers},
journal = {Optics Communications},
volume = {62},
number = {6},
pages = {374-379},
year = {1987},
issn = {0030-4018},
doi = {https://doi.org/10.1016/0030-4018(87)90003-4},
url = {https://www.sciencedirect.com/science/article/pii/0030401887900034},
author = {P. Emplit and J.P. Hamaide and F. Reynaud and C. Froehly and A. Barthelemy}
}

@article{Bossert2023,
author = {Marine Bossert  and I. Trimaille  and L. Cagnon  and B. Chabaud  and C. Gueneau  and P. Spathis  and P. E. Wolf  and E. Rolley },
title = {Surface tension of cavitation bubbles},
journal = {Proceedings of the National Academy of Sciences},
volume = {120},
number = {15},
pages = {e2300499120},
year = {2023},
doi = {10.1073/pnas.2300499120},
URL = {https://www.pnas.org/doi/abs/10.1073/pnas.2300499120}}

@article{Chiao1964,
  title = {Self-Trapping of Optical Beams},
  author = {Chiao, R. Y. and Garmire, E. and Townes, C. H.},
  journal = {Phys. Rev. Lett.},
  volume = {13},
  issue = {15},
  pages = {479--482},
  numpages = {0},
  year = {1964},
  month = {Oct},
  publisher = {American Physical Society},
  doi = {10.1103/PhysRevLett.13.479},
  url = {https://link.aps.org/doi/10.1103/PhysRevLett.13.479}
}

@article{Gammal2001,
  title = {Critical number of atoms for attractive {Bose-Einstein} condensates with cylindrically symmetrical traps},
  author = {Gammal, A. and Frederico, T. and Tomio, Lauro},
  journal = {Phys. Rev. A},
  volume = {64},
  issue = {5},
  pages = {055602},
  numpages = {4},
  year = {2001},
  month = {Oct},
  publisher = {American Physical Society},
  doi = {10.1103/PhysRevA.64.055602},
  url = {https://link.aps.org/doi/10.1103/PhysRevA.64.055602}
}

@article{Salasnich2007,
  title = {Nearly-one-dimensional self-attractive {Bose-Einstein} condensates in optical lattices},
  author = {Salasnich, L. and Cetoli, A. and Malomed, B. A. and Toigo, F.},
  journal = {Phys. Rev. A},
  volume = {75},
  issue = {3},
  pages = {033622},
  numpages = {8},
  year = {2007},
  month = {Mar},
  publisher = {American Physical Society},
  doi = {10.1103/PhysRevA.75.033622},
  url = {https://link.aps.org/doi/10.1103/PhysRevA.75.033622}
}

@article{Strutt1879,
  title={On the capillary phenomena of jets},
  author={J.W. Strutt},
  journal={Proc. R. Soc. Lond.},
  volume={29},
  pages={71-79},
  year={1879}
}

@article{Petrov2015a,
  title = {Quantum Mechanical Stabilization of a Collapsing {Bose-Bose} Mixture},
  author = {Petrov, D. S.},
  journal = {Phys. Rev. Lett.},
  volume = {115},
  issue = {15},
  pages = {155302},
  numpages = {5},
  year = {2015},
  month = {Oct},
  publisher = {American Physical Society},
  doi = {10.1103/PhysRevLett.115.155302},
  url = {http://link.aps.org/doi/10.1103/PhysRevLett.115.155302}
}

@article{Strecker2002,
  title = {Formation and propagation of matter-wave soliton trains},
  author = {Kevin Strecker and Guthrie Partridge and Andrew Truscott and Randall Hulet},
  journal = {Nature},
  volume = {417},
  pages = {150--153},
  year = {1999},
  url = {https://www.nature.com/articles/nature747}
}

@article{Cornish2006,
  title = {Formation of Bright Matter-Wave Solitons during the Collapse of Attractive {Bose-Einstein} Condensates},
  author = {Cornish, Simon L. and Thompson, Sarah T. and Wieman, Carl E.},
  journal = {Phys. Rev. Lett.},
  volume = {96},
  issue = {17},
  pages = {170401},
  numpages = {4},
  year = {2006},
  month = {May},
  publisher = {American Physical Society},
  doi = {10.1103/PhysRevLett.96.170401},
  url = {https://link.aps.org/doi/10.1103/PhysRevLett.96.170401}
}

@article{Lepoutre2016,
  title = {Production of strongly bound $^{39}${K} bright solitons},
  author = {Lepoutre, S. and Fouch\'e, L. and Boiss\'e, A. and Berthet, G. and Salomon, G. and Aspect, A. and Bourdel, T.},
  journal = {Phys. Rev. A},
  volume = {94},
  issue = {5},
  pages = {053626},
  numpages = {5},
  year = {2016},
  month = {Nov},
  publisher = {American Physical Society},
  doi = {10.1103/PhysRevA.94.053626},
  url = {https://link.aps.org/doi/10.1103/PhysRevA.94.053626}
}

@article{Zhang2015,
	author = {Zhang, Xiao-Fei and Han, Wei and Wen, Lin and Zhang, Peng and Dong, Rui-Fang and Chang, Hong and Zhang, Shou-Gang},
	doi = {10.1038/srep08684},
	isbn = {2045-2322},
	journal = {Scientific Reports},
	number = {1},
	pages = {8684},
	title = {Two-component dipolar {Bose-Einstein} condensate in concentrically coupled annular traps},
	url = {https://doi.org/10.1038/srep08684},
	volume = {5},
	year = {2015}}

@article{Saito2003,
  title = {Dynamically Stabilized Bright Solitons in a Two-Dimensional {Bose-Einstein} Condensate},
  author = {Saito, Hiroki and Ueda, Masahito},
  journal = {Phys. Rev. Lett.},
  volume = {90},
  issue = {4},
  pages = {040403},
  numpages = {4},
  year = {2003},
  month = {Jan},
  publisher = {American Physical Society},
  doi = {10.1103/PhysRevLett.90.040403},
  url = {https://link.aps.org/doi/10.1103/PhysRevLett.90.040403}
}

@article{Wang2020,
doi = {10.1088/1367-2630/ab725b},
url = {https://dx.doi.org/10.1088/1367-2630/ab725b},
year = {2020},
month = {mar},
publisher = {IOP Publishing},
volume = {22},
number = {3},
pages = {033006},
author = {Wang, Ya-Jun and Wen, Lin and Chen, Guang-Ping and Zhang, Shou-Gang and Zhang, Xiao-Fei},
title = {Formation, stability, and dynamics of vector bright solitons in a trapless {Bose–Einstein} condensate with spin–orbit coupling},
journal = {New Journal of Physics}
}

@article{Chen2021,
  title = {Observation of Scale Invariance in Two-Dimensional Matter-Wave Townes Solitons},
  author = {Chen, Cheng-An and Hung, Chen-Lung},
  journal = {Phys. Rev. Lett.},
  volume = {127},
  issue = {2},
  pages = {023604},
  numpages = {6},
  year = {2021},
  month = {Jul},
  publisher = {American Physical Society},
  doi = {10.1103/PhysRevLett.127.023604},
  url = {https://link.aps.org/doi/10.1103/PhysRevLett.127.023604}
}

@article{Wen2016,
  title = {Motion of solitons in one-dimensional spin-orbit-coupled {Bose-Einstein} condensates},
  author = {Wen, Lin and Sun, Q. and Chen, Yu and Wang, Deng-Shan and Hu, J. and Chen, H. and Liu, W.-M. and Juzeli\ifmmode \bar{u}\else \={u}\fi{}nas, G. and Malomed, Boris A. and Ji, An-Chun},
  journal = {Phys. Rev. A},
  volume = {94},
  issue = {6},
  pages = {061602},
  numpages = {6},
  year = {2016},
  month = {Dec},
  publisher = {American Physical Society},
  doi = {10.1103/PhysRevA.94.061602},
  url = {https://link.aps.org/doi/10.1103/PhysRevA.94.061602}
}

@article{Barbiero2014,
  title = {Quantum bright solitons in a quasi-one-dimensional optical lattice},
  author = {Barbiero, Luca and Salasnich, Luca},
  journal = {Phys. Rev. A},
  volume = {89},
  issue = {6},
  pages = {063605},
  numpages = {5},
  year = {2014},
  month = {Jun},
  publisher = {American Physical Society},
  doi = {10.1103/PhysRevA.89.063605},
  url = {https://link.aps.org/doi/10.1103/PhysRevA.89.063605}
}

@article{DiCarli2019,
  title = {Excitation Modes of Bright Matter-Wave Solitons},
  author = {Di Carli, Andrea and Colquhoun, Craig D. and Henderson, Grant and Flannigan, Stuart and Oppo, Gian-Luca and Daley, Andrew J. and Kuhr, Stefan and Haller, Elmar},
  journal = {Phys. Rev. Lett.},
  volume = {123},
  issue = {12},
  pages = {123602},
  numpages = {5},
  year = {2019},
  month = {Sep},
  publisher = {American Physical Society},
  doi = {10.1103/PhysRevLett.123.123602},
  url = {https://link.aps.org/doi/10.1103/PhysRevLett.123.123602}
}

@article {Cabrera2018,
	author = {Cabrera, C. R. and Tanzi, L. and Sanz, J. and Naylor, B. and Thomas, P. and Cheiney, P. and Tarruell, L.},
	title = {Quantum liquid droplets in a mixture of {Bose-Einstein} condensates},
	volume = {359},
	number = {6373},
	pages = {301--304},
	year = {2018},
	doi = {10.1126/science.aao5686},
	publisher = {American Association for the Advancement of Science},
	issn = {0036-8075},
	journal = {Science}
}

@article{Semeghini2018,
  title = {Self-Bound Quantum Droplets of Atomic Mixtures in Free Space},
  author = {Semeghini, G. and Ferioli, G. and Masi, L. and Mazzinghi, C. and Wolswijk, L. and Minardi, F. and Modugno, M. and Modugno, G. and Inguscio, M. and Fattori, M.},
  journal = {Phys. Rev. Lett.},
  volume = {120},
  issue = {23},
  pages = {235301},
  numpages = {5},
  year = {2018},
  month = {Jun},
  publisher = {American Physical Society},
  doi = {10.1103/PhysRevLett.120.235301},
  url = {https://link.aps.org/doi/10.1103/PhysRevLett.120.235301}
}

@article{Cavicchioli2025,
  title = {Dynamical Formation of Multiple Quantum Droplets in a {Bose-Bose} Mixture},
  author = {Cavicchioli, L. and Fort, C. and Ancilotto, F. and Modugno, M. and Minardi, F. and Burchianti, A.},
  journal = {Phys. Rev. Lett.},
  volume = {134},
  issue = {9},
  pages = {093401},
  numpages = {6},
  year = {2025},
  month = {Mar},
  publisher = {American Physical Society},
  doi = {10.1103/PhysRevLett.134.093401},
  url = {https://link.aps.org/doi/10.1103/PhysRevLett.134.093401}
}

@article{d2019observation,
  title={Observation of quantum droplets in a heteronuclear bosonic mixture},
  author={D'Errico, C and Burchianti, A and Prevedelli, M and Salasnich, L and Ancilotto, F and Modugno, M and Minardi, F and Fort, C},
  journal={Physical Review Research},
  volume={1},
  number={3},
  pages={033155},
  year={2019},
  publisher={APS},
  url = {https://doi.org/10.1103/PhysRevResearch.1.033155}
}

@article{Guo2021,
  title = {Lee-Huang-Yang effects in the ultracold mixture of $^{23}\mathrm{Na}$ and $^{87}\mathrm{Rb}$ with attractive interspecies interactions},
  author = {Guo, Zhichao and Jia, Fan and Li, Lintao and Ma, Yinfeng and Hutson, Jeremy M. and Cui, Xiaoling and Wang, Dajun},
  journal = {Phys. Rev. Res.},
  volume = {3},
  issue = {3},
  pages = {033247},
  numpages = {8},
  year = {2021},
  month = {Sep},
  publisher = {American Physical Society},
  doi = {10.1103/PhysRevResearch.3.033247},
  url = {https://link.aps.org/doi/10.1103/PhysRevResearch.3.033247}
}

@article{Kadau2016a,
	Author = {Kadau, Holger and Schmitt, Matthias and Wenzel, Matthias and Wink, Clarissa and Maier, Thomas and Ferrier-Barbut, Igor and Pfau, Tilman},
	Date = {2016/02/11/print},
	Day = {11},
	Isbn = {0028-0836},
	Journal = {Nature},
	L3 = {10.1038/nature16485},
	M3 = {Letter},
	Month = {02},
	Number = {7589},
	Pages = {194--197},
	Publisher = {Nature Publishing Group, a division of Macmillan Publishers Limited. All Rights Reserved.},
	Title = {Observing the {Rosensweig} instability of a quantum ferrofluid},
	Ty = {JOUR},
	Url = {http://dx.doi.org/10.1038/nature16485},
	Volume = {530},
	Year = {2016}}

@article{Astrakharchik2018,
  title = {Dynamics of one-dimensional quantum droplets},
  author = {Astrakharchik, G. E. and Malomed, B. A.},
  journal = {Phys. Rev. A},
  volume = {98},
  issue = {1},
  pages = {013631},
  numpages = {11},
  year = {2018},
  month = {Jul},
  publisher = {American Physical Society},
  doi = {10.1103/PhysRevA.98.013631},
  url = {https://link.aps.org/doi/10.1103/PhysRevA.98.013631}
}

@article{Petrov2016,
  title = {Ultradilute Low-Dimensional Liquids},
  author = {Petrov, D. S. and Astrakharchik, G. E.},
  journal = {Phys. Rev. Lett.},
  volume = {117},
  issue = {10},
  pages = {100401},
  numpages = {5},
  year = {2016},
  month = {Sep},
  publisher = {American Physical Society},
  doi = {10.1103/PhysRevLett.117.100401},
  url = {https://link.aps.org/doi/10.1103/PhysRevLett.117.100401}
}

@article{Spada2024,
  title = {Quantum Droplets in Two-Dimensional {Bose} Mixtures at Finite Temperature},
  author = {Spada, G. and Pilati, S. and Giorgini, S.},
  journal = {Phys. Rev. Lett.},
  volume = {133},
  issue = {8},
  pages = {083401},
  numpages = {6},
  year = {2024},
  month = {Aug},
  publisher = {American Physical Society},
  doi = {10.1103/PhysRevLett.133.083401},
  url = {https://link.aps.org/doi/10.1103/PhysRevLett.133.083401}
}

@article{Dong2022,
  title = {Internal modes of two-dimensional quantum droplets},
  author = {Dong, Liangwei and Shi, Kai and Huang, Changming},
  journal = {Phys. Rev. A},
  volume = {106},
  issue = {5},
  pages = {053303},
  numpages = {7},
  year = {2022},
  month = {Nov},
  publisher = {American Physical Society},
  doi = {10.1103/PhysRevA.106.053303},
  url = {https://link.aps.org/doi/10.1103/PhysRevA.106.053303}
}

@article{cheiney2018bright,
  title={Bright soliton to quantum droplet transition in a mixture of {Bose-Einstein} condensates},
  author={Cheiney, P and Cabrera, CR and Sanz, J and Naylor, B and Tanzi, L and Tarruell, L},
  journal={Physical review letters},
  volume={120},
  number={13},
  pages={135301},
  year={2018},
  publisher={APS},
  doi = {10.1103/physrevlett.120.135301}
}

@article{Pathak2022,
title = {Droplet to soliton crossover at negative temperature in presence of bi-periodic optical lattices},
journal = {Scientific Reports},
volume = {12},
pages = {18248},
year = {2022},
url = {https://www.nature.com/articles/s41598-022-23026-x},
author = {M. R. Pathak and A. Nath}
}

@article{cappellaro2018collective,
  title={Collective modes across the soliton-droplet crossover in binary {Bose} mixtures},
  author={Cappellaro, Alberto and Macr{\`\i}, Tommaso and Salasnich, Luca},
  journal={Physical Review A},
  volume={97},
  number={5},
  pages={053623},
  year={2018},
  publisher={APS},
  url = {https://doi.org/10.1103/PhysRevA.97.053623}
}

@article{Baillie2017,
  title = {Collective Excitations of Self-Bound Droplets of a Dipolar Quantum Fluid},
  author = {Baillie, D. and Wilson, R. M. and Blakie, P. B.},
  journal = {Phys. Rev. Lett.},
  volume = {119},
  issue = {25},
  pages = {255302},
  numpages = {5},
  year = {2017},
  month = {Dec},
  publisher = {American Physical Society},
  doi = {10.1103/PhysRevLett.119.255302},
  url = {https://link.aps.org/doi/10.1103/PhysRevLett.119.255302}
}

@article{bisset2021quantum,
  title={Quantum droplets of dipolar mixtures},
  author={Bisset, RN and Ardila, LA Pe{\~n}a and Santos, Luis},
  journal={Physical Review Letters},
  volume={126},
  number={2},
  pages={025301},
  year={2021},
  publisher={APS}
}

@article{edmonds2020quantum,
  title={Quantum droplets of quasi-one-dimensional dipolar {Bose-Einstein} condensates},
  author={M. Edmonds and T. Bland and N. Parker},
  journal={Journal of Physics Communications},
  volume={4},
  number={12},
  pages={125008},
  year={2020},
  publisher={IOP Publishing},
  url = {10.1088/2399-6528/abcc3b}
}

@article{Bottcher2019,
  title = {Transient Supersolid Properties in an Array of Dipolar Quantum Droplets},
  author = {B\"ottcher, Fabian and Schmidt, Jan-Niklas and Wenzel, Matthias and Hertkorn, Jens and Guo, Mingyang and Langen, Tim and Pfau, Tilman},
  journal = {Phys. Rev. X},
  volume = {9},
  issue = {1},
  pages = {011051},
  numpages = {7},
  year = {2019},
  month = {Mar},
  publisher = {American Physical Society},
  doi = {10.1103/PhysRevX.9.011051},
  url = {https://link.aps.org/doi/10.1103/PhysRevX.9.011051}
}

@article{tanzi2019supersolid,
  title={Supersolid symmetry breaking from compressional oscillations in a dipolar quantum gas},
  author={Tanzi, L and Roccuzzo, SM and Lucioni, E and Fam{\`a}, F and Fioretti, A and Gabbanini, C and Modugno, G and Recati, A and Stringari, S},
  journal={Nature},
  volume={574},
  number={7778},
  pages={382--385},
  year={2019},
  publisher={Nature Publishing Group}
}

@article{Chomaz2019,
  title = {Long-Lived and Transient Supersolid Behaviors in Dipolar Quantum Gases},
  author = {Chomaz, L. and Petter, D. and Ilzh\"ofer, P. and Natale, G. and Trautmann, A. and Politi, C. and Durastante, G. and van Bijnen, R. M. W. and Patscheider, A. and Sohmen, M. and Mark, M. J. and Ferlaino, F.},
  journal = {Phys. Rev. X},
  volume = {9},
  issue = {2},
  pages = {021012},
  numpages = {12},
  year = {2019},
  month = {Apr},
  publisher = {American Physical Society},
  doi = {10.1103/PhysRevX.9.021012},
  url = {https://link.aps.org/doi/10.1103/PhysRevX.9.021012}
}

@article{norcia2021two,
   title={Two-dimensional supersolidity in a dipolar quantum gas},
   volume={596},
   ISSN={1476-4687},
   url={http://dx.doi.org/10.1038/s41586-021-03725-7},
   DOI={10.1038/s41586-021-03725-7},
   number={7872},
   journal={Nature},
   publisher={Springer Science and Business Media LLC},
   author={Norcia, Matthew A. and Politi, Claudia and Klaus, Lauritz and Poli, Elena and Sohmen, Maximilian and Mark, Manfred J. and Bisset, Russell N. and Santos, Luis and Ferlaino, Francesca},
   year={2021},
   month={Aug},
   pages={357–361}
}

@article{Gligori2008,
  title = {Bright solitons in the one-dimensional discrete Gross-Pitaevskii equation with dipole-dipole interactions},
  author = {Gligori\ifmmode \acute{c}\else \'{c}\fi{}, Goran and Maluckov, Aleksandra and Had\ifmmode \check{z}\else \v{z}\fi{}ievski, Ljup\ifmmode \check{c}\else \v{c}\fi{}o and Malomed, Boris A.},
  journal = {Phys. Rev. A},
  volume = {78},
  issue = {6},
  pages = {063615},
  numpages = {10},
  year = {2008},
  month = {Dec},
  publisher = {American Physical Society},
  doi = {10.1103/PhysRevA.78.063615},
  url = {https://link.aps.org/doi/10.1103/PhysRevA.78.063615}
}

@article{Lakomy2012,
  title = {Soliton molecules in dipolar {Bose-Einstein} condensates},
  author = {\L{}akomy, Kazimierz and Nath, Rejish and Santos, Luis},
  journal = {Phys. Rev. A},
  volume = {86},
  issue = {1},
  pages = {013610},
  numpages = {7},
  year = {2012},
  month = {Jul},
  publisher = {American Physical Society},
  doi = {10.1103/PhysRevA.86.013610},
  url = {https://link.aps.org/doi/10.1103/PhysRevA.86.013610}
}

@article{Pedri2005a,
  title = {Two-Dimensional Bright Solitons in Dipolar {Bose-Einstein} Condensates},
  author = {Pedri, P. and Santos, L.},
  journal = {Phys. Rev. Lett.},
  volume = {95},
  issue = {20},
  pages = {200404},
  numpages = {4},
  year = {2005},
  month = {Nov},
  publisher = {American Physical Society},
  doi = {10.1103/PhysRevLett.95.200404},
  url = {http://link.aps.org/doi/10.1103/PhysRevLett.95.200404}
}

@article{Huang2017,
  title = {Dipolar bright solitons and solitary vortices in a radial lattice},
  author = {Huang, Chunqing and Lyu, Lin and Huang, Hao and Chen, Zhaopin and Fu, Shenhe and Tan, Haishu and Malomed, Boris A. and Li, Yongyao},
  journal = {Phys. Rev. A},
  volume = {96},
  issue = {5},
  pages = {053617},
  numpages = {9},
  year = {2017},
  month = {Nov},
  publisher = {American Physical Society},
  doi = {10.1103/PhysRevA.96.053617},
  url = {https://link.aps.org/doi/10.1103/PhysRevA.96.053617}
}

@article{Edmonds2017,
  title = {Engineering bright matter-wave solitons of dipolar condensates},
  author = {M. J. Edmonds, T. Bland, R. Doran and N. G. Parker},
  journal = {New J. Phys.},
  volume = {19},
  pages = {023019},
  year = {2017},
  doi = {10.1088/1367-2630/aa5a6b},
  url = {https://iopscience.iop.org/article/10.1088/1367-2630/aa5a6b}
}

@article{Petrov2000,
  title = {Regimes of Quantum Degeneracy in Trapped 1D Gases},
  author = {Petrov, D. S. and Shlyapnikov, G. V. and Walraven, J. T. M.},
  journal = {Phys. Rev. Lett.},
  volume = {85},
  issue = {18},
  pages = {3745--3749},
  numpages = {0},
  year = {2000},
  month = {Oct},
  publisher = {American Physical Society},
  doi = {10.1103/PhysRevLett.85.3745},
  url = {https://link.aps.org/doi/10.1103/PhysRevLett.85.3745}
}

@article{Menotti2002,
  title = {Collective oscillations of a one-dimensional trapped {Bose-Einstein} gas},
  author = {Menotti, Chiara and Stringari, Sandro},
  journal = {Phys. Rev. A},
  volume = {66},
  issue = {4},
  pages = {043610},
  numpages = {6},
  year = {2002},
  month = {Oct},
  publisher = {American Physical Society},
  doi = {10.1103/PhysRevA.66.043610},
  url = {https://link.aps.org/doi/10.1103/PhysRevA.66.043610}
}

@article{natale2022bloch,
  title={Bloch oscillations and matter-wave localization of a dipolar quantum gas in a one-dimensional lattice},
  author={Natale, Gabriele and Bland, Thomas and Gschwendtner, Simon and Lafforgue, Louis and Gr{\"u}n, Daniel S and Patscheider, Alexander and Mark, Manfred J and Ferlaino, Francesca},
  journal={Communications Physics},
  volume={5},
  number={1},
  pages={227},
  year={2022},
  publisher={Nature Publishing Group UK London},
  url = {https://doi.org/10.1038/s42005-022-01009-8}
}

@Book{Pitaevskii16,
  Title                    = {{Bose-Einstein} Condensation and Superfluidity},
  Author                   = {L. Pitaevskii and S. Stringari},
  Publisher                = {Oxford University Press},
  Year                     = {2016}
}

@article{poli2021maintaining,
  title={Maintaining supersolidity in one and two dimensions},
  author={Poli, Elena and Bland, Thomas and Politi, Claudia and Klaus, Lauritz and Norcia, Matthew A and Ferlaino, Francesca and Bisset, Russell N and Santos, Luis},
  journal={Physical Review A},
  volume={104},
  number={6},
  pages={063307},
  year={2021},
  publisher={APS},
url = {https://journals.aps.org/pra/abstract/10.1103/PhysRevA.104.063307}
}

@article{Korteweg1895,
  title={On the Change of Form of Long Waves Advancing in a Rectangular Canal, and on a New Type of Long Stationary Waves.},
  author={D. J. Korteweg and G. de Vries},
  journal={Philosophical Magazine},
  volume={39},
  number={240},
  pages={422-443},
  year={1895}
}

@article{Otajonov2019,
title = {Stationary and dynamical properties of one-dimensional quantum droplets},
journal = {Physics Letters A},
volume = {383},
number = {34},
pages = {125980},
year = {2019},
issn = {0375-9601},
doi = {https://doi.org/10.1016/j.physleta.2019.125980},
url = {https://www.sciencedirect.com/science/article/pii/S0375960119308473},
author = {Sherzod R. Otajonov and Eduard N. Tsoy and Fatkhulla Kh. Abdullaev}
}

@article{Lavoine2021,
  title = {Beyond-mean-field crossover from one dimension to three dimensions in quantum droplets of binary mixtures},
  author = {Lavoine, L. and Bourdel, T.},
  journal = {Phys. Rev. A},
  volume = {103},
  issue = {3},
  pages = {033312},
  numpages = {7},
  year = {2021},
  month = {Mar},
  publisher = {American Physical Society},
  doi = {10.1103/PhysRevA.103.033312},
  url = {https://link.aps.org/doi/10.1103/PhysRevA.103.033312}
}

@article{Stenger1999,
  title = {Bragg Spectroscopy of a {Bose-Einstein} Condensate},
  author = {Stenger, J. and Inouye, S. and Chikkatur, A. P. and Stamper-Kurn, D. M. and Pritchard, D. E. and Ketterle, W.},
  journal = {Phys. Rev. Lett.},
  volume = {82},
  issue = {23},
  pages = {4569--4573},
  numpages = {0},
  year = {1999},
  month = {Jun},
  publisher = {American Physical Society},
  doi = {10.1103/PhysRevLett.82.4569},
  url = {https://link.aps.org/doi/10.1103/PhysRevLett.82.4569}
}

@article{blakie2002theory,
  title={Theory of coherent Bragg spectroscopy of a trapped {Bose-Einstein} condensate},
  author={Blakie, PB and Ballagh, RJ and Gardiner, CW},
  journal={Physical Review A},
  volume={65},
  number={3},
  pages={033602},
  year={2002},
  publisher={APS}
}

@article{Houwman2024,
  title = {Measurement of the Excitation Spectrum of a Dipolar Gas in the Macrodroplet Regime},
  author = {Houwman, J. J. A. and Baillie, D. and Blakie, P. B. and Natale, G. and Ferlaino, F. and Mark, M. J.},
  journal = {Phys. Rev. Lett.},
  volume = {132},
  issue = {10},
  pages = {103401},
  numpages = {6},
  year = {2024},
  month = {Mar},
  publisher = {American Physical Society},
  doi = {10.1103/PhysRevLett.132.103401},
  url = {https://link.aps.org/doi/10.1103/PhysRevLett.132.103401}
}

@article{Pollack2010,
  title = {Collective excitation of a {Bose-Einstein} condensate by modulation of the atomic scattering length},
  author = {Pollack, S. E. and Dries, D. and Hulet, R. G. and Magalh\~aes, K. M. F. and Henn, E. A. L. and Ramos, E. R. F. and Caracanhas, M. A. and Bagnato, V. S.},
  journal = {Phys. Rev. A},
  volume = {81},
  issue = {5},
  pages = {053627},
  numpages = {5},
  year = {2010},
  month = {May},
  publisher = {American Physical Society},
  doi = {10.1103/PhysRevA.81.053627},
  url = {https://link.aps.org/doi/10.1103/PhysRevA.81.053627}
}

@article{Haller2009,
  author       = {E. Haller and M. Gustavsson and M. J. Mark and J. G. Danzl and R. Hart and G. Pupillo and H.-C. N\"agerl},
  title        = {Realization of an Excited, Strongly Correlated Quantum Gas Phase},
  journal      = {Science},
  volume       = {325},
  number       = {5945},
  pages        = {1224--1227},
  year         = {2009},
  doi          = {10.1126/science.1175850},
  publisher    = {American Association for the Advancement of Science},
  issn         = {0036-8075},
  url          = {https://science.sciencemag.org/content/325/5945/1224}
}

@article{guo2019low,
  title={The low-energy Goldstone mode in a trapped dipolar supersolid},
  author={Guo, Mingyang and B{\"o}ttcher, Fabian and Hertkorn, Jens and Schmidt, Jan-Niklas and Wenzel, Matthias and B{\"u}chler, Hans Peter and Langen, Tim and Pfau, Tilman},
  journal={Nature},
  volume={574},
  number={7778},
  pages={386--389},
  year={2019},
  publisher={Nature Publishing Group}
}

@article{ferrier2018scissors,
  title={Scissors mode of dipolar quantum droplets of dysprosium atoms},
  author={Ferrier-Barbut, Igor and Wenzel, Matthias and B{\"o}ttcher, Fabian and Langen, Tim and Isoard, Mathieu and Stringari, Sandro and Pfau, Tilman},
  journal={Physical review letters},
  volume={120},
  number={16},
  pages={160402},
  year={2018},
  publisher={APS},
  url = {https://doi.org/10.1103/PhysRevLett.120.160402}
}

@article{Malkin1991,
title = {Elementary excitations for solitons of the nonlinear Schrödinger equation},
journal = {Physica D: Nonlinear Phenomena},
volume = {53},
number = {1},
pages = {25-32},
year = {1991},
issn = {0167-2789},
doi = {https://doi.org/10.1016/0167-2789(91)90161-2},
url = {https://www.sciencedirect.com/science/article/pii/0167278991901612},
author = {V.M. Malkin and E.G. Shapiro}
}

@article{bakkali2023cross,
  title={The cross-over from Townes solitons to droplets in a 2D {Bose} mixture},
  author={Bakkali-Hassani, B and Maury, C and Stringari, S and Nascimbene, S and Dalibard, J and Beugnon, J},
  journal={New Journal of Physics},
  volume={25},
  number={1},
  pages={013007},
  year={2023},
  publisher={IOP Publishing},
  doi = {10.1088/1367-2630/acaee3}
}

@article{Tikhonenkov2008,
  title = {Anisotropic Solitons in Dipolar {Bose-Einstein} Condensates},
  author = {Tikhonenkov, I. and Malomed, B. A. and Vardi, A.},
  journal = {Phys. Rev. Lett.},
  volume = {100},
  issue = {9},
  pages = {090406},
  numpages = {4},
  year = {2008},
  month = {Mar},
  publisher = {American Physical Society},
  doi = {10.1103/PhysRevLett.100.090406},
  url = {https://link.aps.org/doi/10.1103/PhysRevLett.100.090406}
}

@article{Stringari1996,
  title = {Collective Excitations of a Trapped {Bose}-Condensed Gas},
  author = {Stringari, S.},
  journal = {Phys. Rev. Lett.},
  volume = {77},
  issue = {12},
  pages = {2360--2363},
  numpages = {0},
  year = {1996},
  month = {Sep},
  publisher = {American Physical Society},
  doi = {10.1103/PhysRevLett.77.2360},
  url = {https://link.aps.org/doi/10.1103/PhysRevLett.77.2360}
}

@article{chomaz2016quantum,
  title={Quantum-fluctuation-driven crossover from a dilute {Bose-Einstein} condensate to a macrodroplet in a dipolar quantum fluid},
  author={Chomaz, L and Baier, S and Petter, D and Mark, MJ and W{\"a}chtler, F and Santos, Luis and Ferlaino, F},
  journal={Physical Review X},
  volume={6},
  number={4},
  pages={041039},
  year={2016},
  publisher={APS},
  url = {https://doi.org/10.1103/PhysRevX.6.041039}
}

@article{Bakkali2021,
  title = {Realization of a Townes Soliton in a Two-Component Planar {Bose} Gas},
  author = {Bakkali-Hassani, B. and Maury, C. and Zou, Y.-Q. and Le Cerf, \'E. and Saint-Jalm, R. and Castilho, P. C. M. and Nascimbene, S. and Dalibard, J. and Beugnon, J.},
  journal = {Phys. Rev. Lett.},
  volume = {127},
  issue = {2},
  pages = {023603},
  numpages = {6},
  year = {2021},
  month = {Jul},
  publisher = {American Physical Society},
  doi = {10.1103/PhysRevLett.127.023603},
  url = {https://link.aps.org/doi/10.1103/PhysRevLett.127.023603}
}

@article{Bttcher2021,
doi = {10.1088/1361-6633/abc9ab},
url = {https://doi.org/10.1088/1361-6633/abc9ab},
year = {2020},
month = {dec},
publisher = {IOP Publishing},
volume = {84},
number = {1},
pages = {012403},
author = {Böttcher, Fabian and Schmidt, Jan-Niklas and Hertkorn, Jens and Ng, Kevin S H and Graham, Sean D and Guo, Mingyang and Langen, Tim and Pfau, Tilman},
title = {New states of matter with fine-tuned interactions: quantum droplets and dipolar supersolids},
journal = {Reports on Progress in Physics}
}

@article{Lu2010,
  title = {Spatial density oscillations in trapped dipolar condensates},
  author = {Lu, H.-Y. and Lu, H. and Zhang, J.-N. and Qiu, R.-Z. and Pu, H. and Yi, S.},
  journal = {Phys. Rev. A},
  volume = {82},
  issue = {2},
  pages = {023622},
  numpages = {7},
  year = {2010},
  month = {Aug},
  publisher = {American Physical Society},
  doi = {10.1103/PhysRevA.82.023622},
  url = {https://link.aps.org/doi/10.1103/PhysRevA.82.023622}
}

@article{Bao2003,
title = {Numerical solution of the Gross–Pitaevskii equation for {Bose–Einstein} condensation},
journal = {Journal of Computational Physics},
volume = {187},
number = {1},
pages = {318-342},
year = {2003},
issn = {0021-9991},
doi = {https://doi.org/10.1016/S0021-9991(03)00102-5},
url = {https://www.sciencedirect.com/science/article/pii/S0021999103001025},
author = {Weizhu Bao and Dieter Jaksch and Peter A. Markowich}
}

@article{giovanazzi2006expansion,
  title={Expansion dynamics of a dipolar {Bose-Einstein} condensate},
  author={Giovanazzi, S and Pedri, P and Santos, L and Griesmaier, A and Fattori, M and Koch, T and Stuhler, J and Pfau, T},
  journal={Physical Review A},
  volume={74},
  number={1},
  pages={013621},
  year={2006},
  publisher={APS}
}

@article{Glaum2007A,
  volume = {98},
  journal = {Phys. Rev. Lett.},
  author = {Glaum, Konstantin and Pelster, Axel and Kleinert, Hagen and Pfau, Tilman},
  month = {Feb},
  url = {http://link.aps.org/doi/10.1103/PhysRevLett.98.080407},
  doi = {10.1103/PhysRevLett.98.080407},
  year = {2007},
  title = {Critical Temperature of Weakly Interacting Dipolar Condensates},
  issue = {8},
  publisher = {American Physical Society},
  numpages = {4},
  pages = {080407}
}

@Article{Blakie2023,
AUTHOR = {Blakie, Peter Blair},
TITLE = {Axial Collective Mode of a Dipolar Quantum Droplet},
JOURNAL = {Photonics},
VOLUME = {10},
YEAR = {2023},
NUMBER = {4},
ARTICLE-NUMBER = {393},
URL = {https://www.mdpi.com/2304-6732/10/4/393},
ISSN = {2304-6732},
DOI = {10.3390/photonics10040393}
}

@article{Tylutki2020,
  title = {Collective excitations of a one-dimensional quantum droplet},
  author = {Tylutki, Marek and Astrakharchik, Grigori E. and Malomed, Boris A. and Petrov, Dmitry S.},
  journal = {Phys. Rev. A},
  volume = {101},
  issue = {5},
  pages = {051601},
  numpages = {5},
  year = {2020},
  month = {May},
  publisher = {American Physical Society},
  doi = {10.1103/PhysRevA.101.051601},
  url = {https://link.aps.org/doi/10.1103/PhysRevA.101.051601}
}

\end{document}